\documentclass[12pt,preprint]{aastex}

\usepackage{wasysym}
\usepackage{emulateapj5}
\usepackage{textcomp}

\slugcomment{Accepted in ApJ}

\shorttitle{Isotope Heterogeneity of the Protosolar Nebula}
\shortauthors{Dauphas et al.}

\begin{document}


\title{Molybdenum Evidence for Inherited Planetary Scale Isotope\\ Heterogeneity of the Protosolar Nebula}


\author{N. Dauphas, B. Marty\altaffilmark{1}, and L. Reisberg}
\affil{Centre de Recherches P\'etrographiques et G\'eochimiques, CNRS UPR 2300, 15 rue Notre-Dame des Pauvres, BP 20, 54501 Vand\oe uvre-l\`es-Nancy Cedex France}
\email{dauphas@crpg.cnrs-nancy.fr}


\altaffiltext{1}{\'Ecole Nationale Sup\'erieure de G\'eologie, rue du doyen Marcel Roubault, BP 40, 54501 Vand\oe uvre-l\`es-Nancy Cedex France}


\begin{abstract}
 Isotope anomalies provide important information about early solar system evolution. Here we report molybdenum isotope abundances determined in samples of various meteorite classes. There is no fractionation of molybdenum isotopes in our sample set within 0.1 $\permil$ and no contribution from the extinct radionuclide $^{97}$Tc at mass 97 (${\rm ^{97}Tc/^{92}Mo<3\times 10^{-6}}$). Instead, we observe clear anomalies in bulk iron meteorites, mesosiderites, pallasites, and chondrites characterized by a coupled excess in {\it p}- and {\it r}- or a mirror deficit in {\it s}-process nuclides (Mo-HL).  This large scale isotope heterogeneity of the solar system observed for molybdenum must have been inherited from the interstellar environment where the sun was born, illustrating the concept of ``cosmic chemical memory''. The presence of molybdenum anomalies is used to discuss the filiation between planetesimals.
\end{abstract}




\section{Introduction}
Variations of isotope abundances within the solar system that depart from mass-dependent fractionation and nuclear effects may occur as a result of inheritance of presolar nucleosynthetic anomalies \citep{anders93} and mass-independent isotope effects \citep{thiemens99}. Such variations have been detected in bulk meteorite samples for a variety of elements including oxygen \citep{clayton93}, sulfur \citep{thiemens95,farquhar00}, titanium \citep{niemeyer88}, chromium \citep{shukolyukov98,podosek99}, zirconium \citep{yin01}, molybdenum \citep{yin00}, and barium \citep{harper92}, although some of these observations need confirmation \citep{thiemens95,farquhar00,yin01,yin00,harper92}. These variations bear information on important subjects such as stellar nucleosynthesis, galactic chemical evolution, heterogeneity of the protosolar nebula, and filiation of planetary objects. 

Molybdenum is a promising element to use to address most of these points. Indeed, unprocessed presolar carbide and graphite grains carry extreme molybdenum signatures inherited from the stellar environment where they formed \citep{nicolussi98a,nicolussi98b,pellin99,pellin00} and subtle nucleosynthetic anomalies have been detected in macroscopic meteorite samples \citep{yin00}. In addition, the $p$-process nuclide $^{97}$Tc which decays by electron capture to $^{97}$Mo with a mean life of 3.8 Ma may have been alive in the early solar system \citep{yin98}, raising the possibility that this element can be used as an extinct chronometer \citep{podosek96}. Molybdenum is a highly refractory element \citep{fegley93}, so it has largely avoided exchange between gas and dust in the forming solar system. Furthermore, it is a moderately siderophile element \citep{schmitt89} and therefore is relatively abundant in all meteorite classes, so that possible anomalies could be used to trace the filiation between planetesimals.

\section{Mo-HL Isotope Anomalies}

A protocol based on solvent extraction, ion exchange, and plasma ionization mass spectrometry was developed that permits the precise and accurate determination of molybdenum isotope abundances in natural samples \citep{dauphas01}. The method takes advantage of the affinity of molybdenum for di(2-ethylhexyl) phosphate and AG1-X8 strongly basic anion exchanger in order to achieve fine separation of  this element from interfering species  \citep{dauphas01,qilu92a,qilu92b,qilu94}. Isotopic analyses were performed using a Micromass Isoprobe plasma ionization mass spectrometer. Molybdenum isotope abundances were corrected for zirconium \citep{nomura83} and ruthenium \citep{huang97} isobaric interferences by monitoring the ion beam signal at masses 91 and 99. As we were mostly interested in non mass-dependent variations, we employed an internal normalization procedure (${\rm ^{98}Mo/^{96}Mo=1.4470}$) in order to correct molybdenum isotope abundances for both natural and instrumental mass fractionation \citep{dauphas01} using the exponential law \citep{marechal99}. After internal normalization, no variation is observed among natural terrestrial samples, which demonstrates the reliability of the method \citep{dauphas01}. Molybdenum and rhenium concentrations were determined by use of the standard addition technique. Molybdenum isotopic composition is expressed in $\epsilon$-units, which are relative deviations of the sample relative to terrestrial molybdenum isotopic composition in parts per $10^{4}$ (\textpertenthousand), 
\begin{displaymath}
{\rm \epsilon ^{\it i}=[(^{\it i}Mo/^{96}Mo)/(^{\it i}Mo/^{96}Mo)_{std}-1]\times 10^{4}}, 
\end{displaymath}
where $i=92$, 94, 95, 96, 97, 98, or 100. We monitored the raw ratios for mass fractionation by comparison with an external standard but found no variation in our sample set within $\sim 0.1$ ${\rm \permil /amu}$ (Table 1).

The Allende CV3.2 chondrite is a Rosetta stone for deciphering early solar system evolution. We confirm (Fig. 1a) the presence of molybdenum nucleosynthetic anomalies in macroscopic samples of this meteorite \citep{yin00,yin00b}. As illustrated in Fig. 1b, the anomalous isotopic composition observed in Allende may result either from a coupled enrichment in $p$- and $r$- or a mirror deficit in $s$-nuclides \citep{anders89,burbidge57,cameron57}. It has been suggested that the molybdenum composition of Allende resulted from incomplete digestion of presolar grains that bear isotopic signatures typical of $s$-nucleosynthesis \citep{yin00}. Simple calculation \citep{becker87,huss95,nicolussi98a,nicolussi98b,dauphas01} indicates that this cannot be the case because the resulting anomalies would be more than two orders of magnitude smaller than those observed \citep{yin00b}. This implies that anomalous molybdenum in bulk Allende must be carried by another phase that needs to be identified, possibly metal, sulfide, or oxide presolar grains. In reference to Xe-HL \citep{anders93,reynolds64,manuel72}, we denote the coupled anomalies in $p$- and $r$-nuclides Mo-HL. 

Differentiation processes homogenize isotopic compositions originally existing within parent bodies. Thus, differentiated meteorites provide a means of examining isotopic heterogeneity in the protosolar nebula at scales comparable to the regions sampled by the parent bodies, which had masses inferred from metallographic cooling rates to be $10^{16}-10^{19}$ kg \citep{wood64,goldstein65,mittlefehldt98}. For this reason, we measured molybdenum isotope abundances in iron meteorites, mesosiderites, and pallasites (Fig. 1a, top right). Most meteorite groups exhibit similar anomalies to those observed in Allende but of lesser magnitude. Because we observe no decoupling within uncertainties between the $p$- and $r$-anomalies, we cannot decide whether they result from the presence of $p$-, $r$-  or the absence of $s$-presolar phases. In either case, the observed anomalies provide evidence for large scale inherited isotope heterogeneity of the protosolar nebula.  

The stellar environments where $p$-, $r$-, and $s$-nucleosynthesis develop are not identical \citep{burbidge57,cameron57} and the host phases for these signatures are variable \citep{anders93} so that the interstellar medium must be chemically and isotopically heterogeneous. Thus, molybdenum isotope abundances were heterogeneously distributed in the solar system parental molecular cloud and the large scale variations we observe were inherited from the interstellar environment where the sun was born, illustrating the concept of ``cosmic chemical memory'' \citep{clayton82}. 

It was suggested that there might be a heterogeneous distribution of short-lived nuclides ${\rm ^{53}Mn}$ and ${\rm ^{182}Hf}$ in the early solar system \citep{lugmair98,lee00}. Instead, the discovery of inherited isotope anomalies for molybdenum may support some form of ``cosmic chemical memory'' for stable isotopic ratios  ${\rm ^{53}Cr/^{52}Cr}$ and ${\rm ^{182}W/^{184}W}$.

\section{Live $^{97}$Tc in the Early Solar System?}

The positive anomaly at mass $^{97}$Mo could potentially represent a radiogenic contribution from the decay of $^{97}$Tc ($\tau= 3.8$ Ma) as well as a nucleosynthetic component. The ${\rm ^{187}Re-^{187}Os}$ system ($\tau= 62.8$ Ga) offers tempting evidence for diachronism of iron meteorite formation but the inferred chronologies are contradictory \citep{shen96,smoliar96}. The timescale inferred from the ${\rm ^{53}Mn-^{53}Cr}$ system ($\tau= 5.4$ Ma) is equivocal because these nuclides may have been redistributed during the extended cooling history of planetesimals. Yet, some iron meteorites have closure ages within $\sim 7$ Ma of Allende refractory inclusions \citep{hutcheon92}. The time span between metal condensation in the protosolar nebula and core crystallization in asteroids is best estimated from the ${\rm ^{107}Pd-^{107}Ag}$ system ($\tau=9.4$ Ma) to be lower than 10 Ma \citep{chen96}. Thus, iron meteorites crystallized from a metallic magma within $\sim 10$ Ma of the solar system birth, when $^{97}$Tc might have still been alive. Furthermore, the solid/liquid metal partition coefficient of rhenium \citep{fleet99} and by inference technetium is higher than that of molybdenum \citep{liu01}, thus the Re(Tc)/Mo ratio \citep{yin98} should have been extensively fractionated during metal crystallization (Table 1). The fractionation factor notation \citep{jacobsen84} is employed for the Re(Tc)/Mo ratio,
\begin{displaymath}
{\rm {\it f}_{Re/Mo}=(^{185}Re/^{96}Mo)/(^{185}Re/^{96}Mo)_{std}-1},
\end{displaymath} 
where the standard ratio is that of the chondritic uniform reservoir (chur), ${\rm ^{185}Re/^{96}Mo=0.0454}$ \citep{anders89}. To test whether a significant radiogenic contribution exists on the $^{97}$Mo peak, we use the fact that both $^{97}$Mo and $^{100}$Mo are $r$-nuclides, so there must be a relationship between the nucleosynthetic component at mass 97 and the anomaly at mass 100. Thus, we have corrected molybdenum isotope measurements for the nucleosynthetic contribution at mass 97 by using the observed anomaly at mass 100,
\begin{displaymath}
\epsilon ^{97*}=\epsilon ^{97}-(\rho ^{97}-0.5\rho ^{98})/(1-2\rho ^{98})\times \epsilon ^{100},
\end{displaymath}
where ${\rm \rho^{\it i}=(^{\it i}Mo/^{100}Mo)_{\it r}/(^{\it i}Mo/^{100}Mo)_{\oplus}}$, $r$ and $\oplus$ denote the $r$-process and the Earth, respectively. The term $\rho ^{98}$ enters into the equation because ${\rm ^{98}Mo/^{96}Mo}$ is used to correct measurements for mass fractionation \citep{dauphas01}. It is estimated that $\rho ^{97}=0.43$ and $\rho ^{98}=0.25$, so $\epsilon ^{97*}=\epsilon ^{97}-0.61\times \epsilon^{100}$ \citep{arlandini99}. Rhenium is used as a proxy for technetium since these elements are likely to have very similar behavior during metal crystallization \citep{yin98}. If there were live $^{97}$Tc when the iron meteorites formed and if all iron meteorites crystallized simultaneously \citep{chen96}, there should be a linear relationship \citep{jacobsen84} between the radiogenic contribution at mass 97 ($\epsilon ^{97*}$) and the Re(Tc)/Mo fractionation factor ($f_{\rm Re/Mo}$), the slope of which depends on the ${\rm ^{97}Tc/^{185}Re}$ ratio at the time of closure,
\begin{displaymath}
{\rm \epsilon^{97*}={\it Q}\,(^{97}Tc/^{185}Re)_0\times {\it f}_{Re/Mo}},
\end{displaymath}
where ${\rm {\it Q}=10^4\times (^{185}Re/^{97}Mo)_{chur}}$, $Q\sim 791$ \citep{anders89}. It follows from the definition of $\epsilon ^{97*}$ that this relationship holds only if the Earth evolved with a chondritic ${\rm Tc/Mo}$ while ${\rm ^{97}Tc}$ was still alive. This will be true if the core formed after the extinction of ${\rm ^{97}Tc}$, which seems to be the case \citep{allegre95,galer96,lee95,halliday96}. 

There is no correlation between $f_{\rm Re/Mo}$ and $\epsilon ^{97*}$ which indicates that $^{97}$Tc was extinct when the iron meteorites formed (${\rm ^{97}Tc/^{92}Mo<3\times 10^{-6}}$, ${\rm ^{92}Mo}$ is used for normalization because both are $p$-nuclides and were synthetized in the same stellar environment). This result is consistent with information retrieved from modelling of the chemical evolution of the Galaxy in the solar neighborhood (in preparation).

\section{Inference on Planetary Genetics}

Because there is no radiogenic contribution on $^{97}$Mo and because $\epsilon ^{97}$ is measured much more precisely and accurately than other isotopic ratios,  we used $\epsilon ^{97}$  to trace the filiation between parent bodies (Fig. 2).  Terrestrial rocks (stream sediments, porphyry copper millhead, and synthetic glass) are within 0.05 {\textpertenthousand} of the terrestrial standard value \citep{dauphas01}, which demonstrates that the analytical procedure is accurate. The molybdenum isotopic composition of the silicate Earth is representative of the bulk Earth value because the mantle molybdenum content is largely dominated by the component derived from the proto-Earth rather than from a late accreting veneer \citep{righter97}. Within each meteorite class, measurements on various specimens are consistent with derivation from a common source, which reinforces the genetic significance of the chemical classification \citep{scott75,sears88}. On the basis of oxygen \citep{clayton93} and chromium \citep{shukolyukov98,podosek99,shukolyukov01} isotopic ratios, a genetic relationship between Eagle Station pallasites and some carbonaceous chondrites (ESPAL/CV) has been inferred. This link is corroborated by molybdenum isotope measurements, which demonstrates the virtue of this element for tracing the relationships between planetsimals.  The isotopic data support the proposed genetic associations between type IIIAB iron meteorites, mesosiderites, and main group pallasites (IIIAB/MES/MGPAL) but do not substantiate the link between types IAB and IIICD iron meteorites (IAB/IIICD) \citep{mittlefehldt98}. 

Further analyses must be conducted on both macroscopic samples and separated phases in order to confirm and extend molybdenum isotope taxonomy and search for the presolar carriers of Mo-HL.

 \acknowledgments

We thank M. Denise and G. Kurat for providing us with the samples; F. Robert for insightful comments on the manuscript; and C. Zimmermann and D. Yeghicheyan for analytical support. This work was funded by grants from the PNP (CNES/INSU) and PRISMS SMT4-CT98-2220 (EC). This is contribution xxxx of the Centre de Recherches P\'etrographiques et G\'eochimiques.

\clearpage
\begin{figure}
\epsscale{0.6}
\plotone{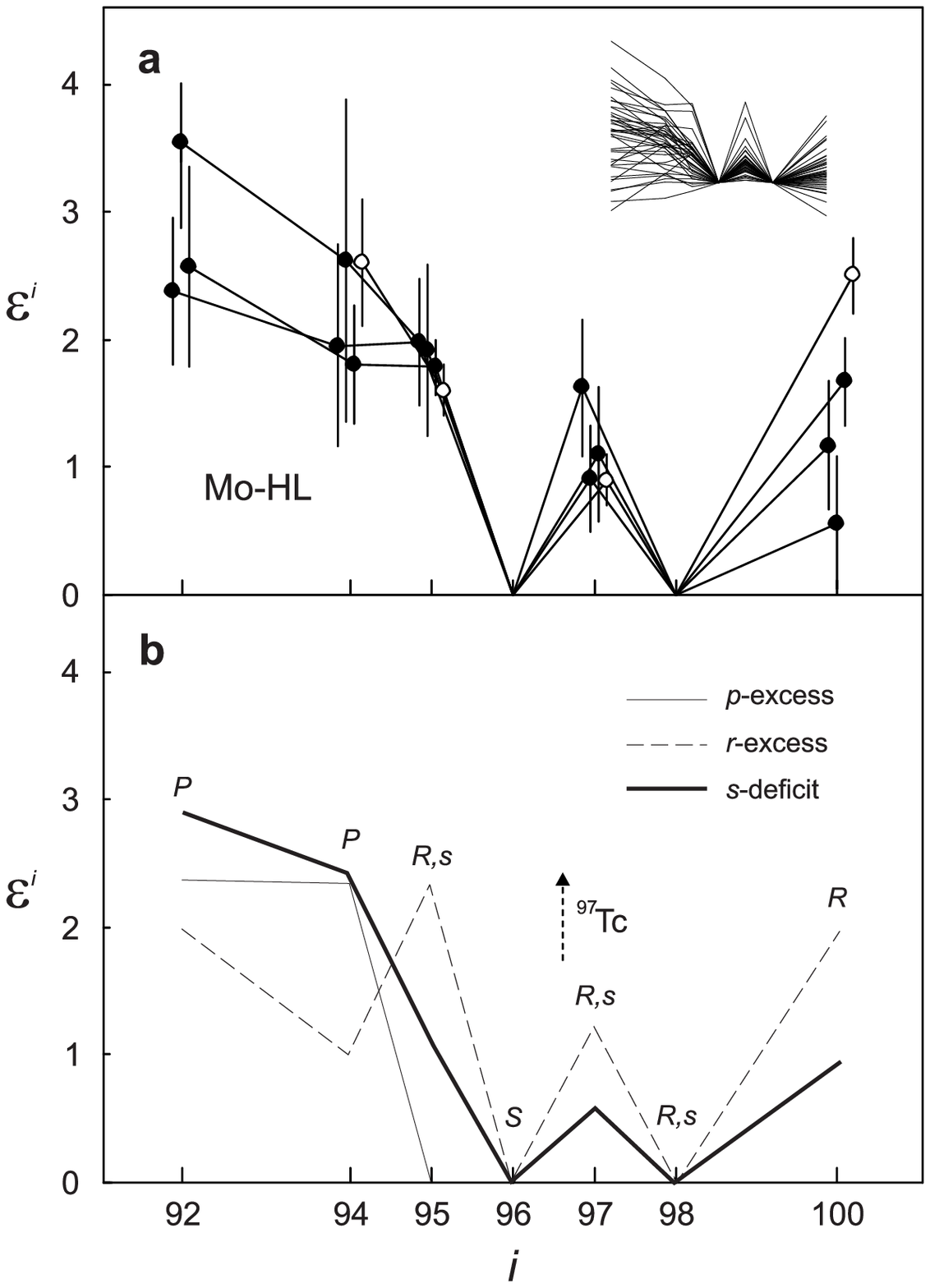}
\caption{Molybdenum isotopic spectra relative to the Earth. a. Molybdenum isotope abundances in Allende  [black (this study) and white \citep{yin00} dots] depart from usual mass-dependent fractionation. Similar anomalies are observed in iron meteorites, mesosiderites, and pallasites (top right) which provides evidence for large scale isotopic heterogeneity of the protosolar nebula. The top curves of the inset are those of Allende. The isotopic spectrum we observe is denoted Mo-HL in reference to Xe-HL \citep{anders93,reynolds64,manuel72}. Uncertainties are 2$\sigma$. b. Synthetic spectra obtained by adding or substracting pure $p$, $r$, and $s$ nucleosynthetic abundances \citep{arlandini99}. The nucleosynthetic sources \citep{burbidge57,cameron57,anders89} of the different nuclides are indicated. There is a possible contribution at mass 97 from the decay of now extinct ${\rm ^{97}Tc}$ ($\tau=3.8$ Ma). The spectrum closest to observations is that corresponding to a deficit in $s$- or a mirror excess in $p$- and $r$- nuclides. \label{fig1}}
\end{figure}

\clearpage 
\begin{figure}
\epsscale{0.5}
\plotone{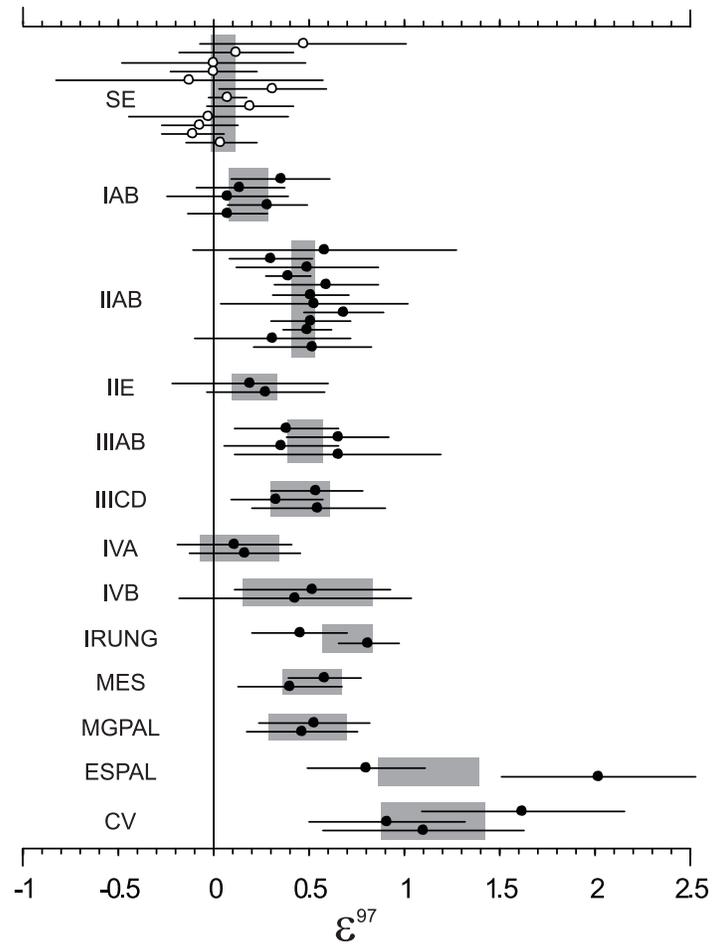}
\caption{Molybdenum isotope taxonomy. SE-silicate Earth \citep{dauphas01}. IAB/IRUNG-iron meteorite. MES-mesosiderite. MGPAL/ESPAL-pallasite. CV-carbonaceous chondrite. The shaded rectangles are the mean composition for each group. Uncertainties are 2$\sigma$. \label{fig2}}
\end{figure}

\clearpage 
\begin{deluxetable}{lccccccccccc}
\tablecolumns{12}
\tablewidth{0pc}
\rotate
\tabletypesize{\tiny}
\tablecaption{Molybdenum Isotope Measurements}
\tablehead{
\colhead{\#} & \colhead{Class} & \colhead{[Mo]}   & \colhead{$f_{\rm Re/Mo}$} & \multicolumn{7}{c}{$\epsilon ^i$ (\textpertenthousand)} & \colhead{$\delta$} \\
\cline{5-11} \\
\colhead{} & \colhead{} & \colhead{ppm} & \colhead{} & \colhead{92} & \colhead{94} & \colhead{95} & \colhead{96} & \colhead{97} & \colhead{98} & \colhead{100} &  \colhead{$\permil$/amu}
}
\startdata
Bitburg & IAB & $6.9\pm 0.7$ & $<-0.8$ & $1.20\pm 1.62$ & $0.34\pm 0.81$ & $0.15\pm 0.40$ & 0 & $0.35\pm 0.26$ & 0 & $-0.49\pm 0.77$ & $-0.01\pm 0.16$ \\
Bitburg & IAB & $6.9\pm 0.7$ & $<-0.8$ & $0.15\pm 0.72$ & $0.22\pm 0.42$ & $-0.10\pm 0.32$ & 0 & $0.14\pm 0.23$ & 0 & $-0.11\pm 0.22$ & $-0.04\pm 0.21$ \\
Canyon Diablo & IAB & $5.1\pm 0.8$ & $-0.05\pm 0.04$ & $-0.48\pm 0.90$ & $-0.39\pm 0.49$ & $-0.21\pm 0.54$ & 0 & $0.07\pm 0.32$ & 0 & $-0.17\pm 0.28$ & $0.34\pm 0.17$ \\
Canyon Diablo & IAB & $5.1\pm 0.8$ & $-0.05\pm 0.04$ & $0.59\pm 0.82$ & $0.45\pm 0.58$ & $0.33\pm 0.38$ & 0 & $0.28\pm 0.21$ & 0 & $0.21\pm 0.44$ & $0.08\pm 0.16$ \\
Canyon Diablo & IAB & $5.1\pm 0.8$ & $-0.05\pm 0.04$ & $-0.19\pm 1.35$ & $0.01\pm 0.45$ & $0.00\pm 0.33$ & 0 & $0.07\pm 0.21$ & 0 & $-0.28\pm0.64$ & $0.12\pm 0.27$ \\
Braunau & IIAB & $7.6\pm 0.1$ & $1.05\pm 0.11$ & $2.50\pm 0.93$ & $1.54\pm 0.71$ & $0.28\pm 0.78$ & 0 & $0.58\pm 0.69$ & 0 & $0.29\pm 0.83$ & $0.09\pm 0.16$ \\
Braunau & IIAB & $7.6\pm 0.1$ & $1.05\pm 0.11$ & $1.53\pm 0.77$ & $1.90\pm 0.83$ & $0.83\pm 0.40$ & 0 & $0.30\pm 0.22$ & 0 & $0.26\pm 0.41$ & $0.04\pm 0.17$ \\
Coahuila & IIAB & $7.5\pm 0.5$ & $2.39\pm0.18$ & $2.15\pm 0.92$ & $1.11\pm 0.27$ & $0.39\pm 0.33$ & 0 & $0.49\pm 0.37$ & 0 & $0.39\pm 0.26$ & $-0.17\pm 0.16$ \\
Coahuila & IIAB & $7.5\pm 0.5$ & $2.39\pm0.18$ & $1.29\pm 0.75$ & $1.41\pm 0.99$ & $0.71\pm 0.37$ & 0 & $0.39\pm 0.12$ & 0 & $0.51\pm 0.16$ & $0.00\pm 0.16$ \\
Coahuila & IIAB & $7.5\pm 0.5$ & $2.39\pm0.18$ & $0.77\pm 1.79$ & $0.70\pm 1.43$ & $0.74\pm 0.74$ & 0 & $0.59\pm 0.27$ & 0 & $0.46\pm 0.55$ & $-0.01\pm 0.16$ \\
Guadalupe Y Calvo & IIAB & $7.0\pm 0.4$ & $6.41\pm 0.37$ & $1.56\pm 0.39$ & $1.48\pm 0.56$ & $0.69\pm 0.28$ & 0 & $0.51\pm 0.20$ & 0 & $0.25\pm 0.22$ & $-0.05\pm 0.17$ \\
Guadalupe Y Calvo & IIAB & $7.0\pm 0.4$ & $6.41\pm 0.37$ & $0.03\pm 0.87$ & $0.93\pm 0.88$ & $0.47\pm 0.46$ & 0 & $0.53\pm 0.49$ & 0 & $0.08\pm 0.70$ & $-0.23\pm 0.45$ \\
Guadalupe Y Calvo & IIAB & $7.0\pm 0.4$ & $6.41\pm 0.37$ & $1.31\pm 0.60$ & $1.19\pm 0.48$ & $0.66\pm 0.39$ & 0 & $0.68\pm 0.21$ & 0 & $0.15\pm 0.26$ & $-0.02\pm 0.18$ \\
Scottsville & IIAB & $6.8\pm 0.8$ & $8.46\pm 0.47$ & $1.57\pm 0.62$ & $1.06\pm 0.57$ & $0.43\pm 0.23$ & 0 & $0.51\pm 0.21$ & 0 & $0.06\pm 0.26$ & $0.06\pm 0.19$ \\
Scottsville & IIAB & $6.8\pm 0.8$ & $8.46\pm 0.47$ & $2.07\pm 0.85$ & $1.59\pm 0.50$ & $0.83\pm 0.25$ & 0 & $0.49\pm 0.13$ & 0 & $0.23\pm 0.30$ & $0.02\pm 0.16$ \\
Sikhote-Alin & IIAB & $5.9\pm 0.5$ & $<-0.8$ & $1.15\pm 1.10$ & $0.86\pm 0.85$ & $0.30\pm 0.64$ & 0 & $0.31\pm 0.41$ & 0 & $0.48\pm 0.44$ & $-0.07\pm 0.20$ \\
Sikhote-Alin & IIAB & $5.9\pm 0.5$ & $<-0.8$ & $1.66\pm 1.23$ & $1.34\pm 0.35$ & $0.82\pm 0.36$ & 0 & $0.52\pm 0.31$ & 0 & $0.23\pm 0.66$ & $0.08\pm 0.24$ \\
Mont Dieu & IIE & $6.8\pm 0.4$ & $-0.02\pm 0.05$ & $1.47\pm 0.64$ & $1.18\pm 0.44$ & $0.66\pm 0.29$ & 0 & $0.19\pm 0.15$ & 0 & $-0.18\pm 0.41$ & $-0.01\pm 0.16$ \\
Mont Dieu & IIE & $6.8\pm 0.4$ & $-0.02\pm 0.05$ & $1.21\pm 0.89$ & $0.91\pm 0.42$ & $0.97\pm 0.26$ & 0 & $0.27\pm 0.19$ & 0 & $-0.84\pm 0.48$ & $-0.28\pm 0.39$ \\
Cape York & IIIAB & $6.3\pm 0.1$ & $0.32\pm 0.07$ & $1.34\pm 0.54$ & $1.00\pm 0.54$ & $0.45\pm 0.15$ & 0 & $0.38\pm 0.27$ & 0 & $0.16\pm 0.30$ & $0.06\pm 0.17$ \\
Cape York & IIIAB & $6.3\pm 0.1$ & $0.32\pm 0.07$ & $2.87\pm 0.76$ & $1.81\pm 0.52$ & $0.88\pm 0.36$ & 0 & $0.65\pm 0.27$ & 0 & $-0.26\pm 0.26$ & $-0.05\pm 0.16$ \\
Henbury & IIIAB & $8.6\pm 2.4$ & $2.96\pm 0.20$ & $1.70\pm 1.55$ & $0.88\pm 0.59$ & $0.09\pm 0.54$ & 0 & $0.35\pm 0.30$ & 0 & $0.02\pm 0.81$ & $0.07\pm 0.17$ \\
Henbury & IIIAB & $8.6\pm 2.4$ & $2.96\pm 0.20$ & $0.81\pm 2.11$ & $0.78\pm 0.77$ & $0.22\pm 1.08$ & 0 & $0.65\pm 0.54$ & 0 & $1.07\pm 1.17$ & $0.07\pm 0.18$ \\
Magnesia & IIICD & $7.4\pm 0.9$ & $<-0.8$ & $1.88\pm 0.61$ & $1.28\pm 0.48$ & $0.58\pm 0.31$ & 0 & $0.54\pm 0.24$ & 0 & $0.01\pm 0.33$ & $0.08\pm 0.15$ \\
Magnesia & IIICD & $7.4\pm 0.9$ & $<-0.8$ & $-0.70\pm 1.92$ & $0.22\pm 0.96$ & $0.18\pm0.38$ & 0 & $0.33\pm 0.24$ & 0 & $-0.06\pm 0.91$ & $-0.25\pm 0.31$ \\
Magnesia & IIICD & $7.4\pm 0.9$ & $<-0.8$ & $1.32\pm 1.41$ & $1.30\pm 0.98$ & $0.74\pm0.50$ & 0 & $0.55\pm 0.35$ & 0 & $0.11\pm 0.56$ & $0.04\pm 0.16$ \\
Gibeon & IVA & $5.7\pm 0.1$ & $-0.27\pm 0.04$ & $0.84\pm 0.71$ & $0.97\pm 0.77$ & $0.75\pm 0.25$ & 0 & $0.11\pm 0.30$ & 0 & $0.21\pm 0.36$ & $0.01\pm 0.17$ \\
Gibeon & IVA & $5.7\pm 0.1$ & $-0.27\pm 0.04$ & $0.44\pm 1.28$ & $0.50\pm 0.78$ & $0.70\pm 0.32$ & 0 & $0.16\pm 0.29$ & 0 & $-0.04\pm 0.26$ & $0.01\pm 0.19$ \\
Cape of Good Hope & IVB & $25.0\pm 0.2$ & $1.58\pm 0.13$ & $2.05\pm 1.26$ & $1.80\pm 0.60$ & $0.86\pm 0.59$ & 0 & $0.52\pm 0.41$ & 0 & $0.38\pm 0.65$ & $0.06\pm 0.17$ \\
Cape of Good Hope & IVB & $25.0\pm 0.2$ & $1.58\pm 0.13$ & $1.75\pm 0.49$ & $1.21\pm 0.16$ & $0.99\pm 0.24$ & 0 & $0.43\pm 0.61$ & 0 & $0.68\pm 1.11$ & $0.06\pm 0.34$ \\
Grand Rapids & IRUNG & $11.5\pm 0.1$ & $1.07\pm 0.10$ & $1.28\pm 0.62$ & $1.15\pm 0.40$ & $0.89\pm 0.25$ & 0 & $0.45\pm 0.25$ & 0 & $0.48\pm 0.36$ & $0.09\pm 0.17$ \\
Grand Rapids & IRUNG & $11.5\pm 0.1$ & $1.07\pm 0.10$ & $0.40\pm 0.70$ & $0.21\pm 0.68$ & $0.73\pm 0.24$ & 0 & $0.81\pm 0.16$ & 0 & $0.60\pm 0.31$ & $-0.04\pm 0.16$ \\
Estherville & MES & \nodata & \nodata & $1.90\pm 0.62$ & $1.64\pm 0.55$ & $1.00\pm 0.30$ & 0 & $0.58\pm 0.19$ & 0 & $0.17\pm 0.50$ & $-0.03\pm 0.25$ \\
Estherville & MES & \nodata & \nodata & $0.52\pm1.42$ & $0.98\pm 0.83$ & $0.60\pm 0.56$ & 0 & $0.40\pm 0.27$ & 0 & $0.30\pm 0.48$ & $-0.18\pm 0.23$ \\
Imilac & MGPAL & \nodata & \nodata & $-0.22\pm 1.97$ & $0.09\pm 1.60$ & $0.09\pm 0.74$ & 0 & $0.53\pm 0.29$ & 0 & $0.67\pm 0.79$ & $-0.05\pm 0.17$ \\
Imilac & MGPAL & \nodata & \nodata & $0.08\pm 1.58$ & $0.18\pm 1.26$ & $-0.04\pm 0.73$ & 0 & $0.46\pm 0.29$ & 0 & $1.10\pm 0.48$ & $-0.12\pm 0.18$ \\
Eagle Sation & ESPAL & \nodata & \nodata & $1.57\pm 0.42$ & $1.21\pm 0.53$ & $0.86\pm 0.35$ & 0 & $0.80\pm 0.31$ & 0 & $0.84\pm 0.30$ & $-0.03\pm 0.16$ \\
Eagle Sation & ESPAL & \nodata & \nodata & $0.43\pm 1.55$ & $1.43\pm 1.24$ & $1.35\pm 0.55$ & 0 & $2.02\pm 0.51$ & 0 & $1.54\pm 0.68$ & $0.03\pm 0.17$ \\
Allende & CV &\nodata  & \nodata & $2.38\pm 0.57$ & $1.95\pm 0.79$ & $1.98\pm 0.50$ & 0 & $1.62\pm 0.53$ & 0 & $1.17\pm 0.50$ & $-0.06\pm 0.38$ \\
Allende & CV & \nodata &\nodata & $3.54\pm 0.67$ & $2.62\pm 1.26$ & $1.92\pm 0.67$ & 0 & $0.91\pm 0.41$ & 0 & $0.56\pm 0.52$ & $0.17\pm 0.23$ \\
Allende & CV & \nodata & \nodata& $2.57\pm 0.78$ & $1.80\pm 0.46$ & $1.78\pm 0.21$ & 0 & $1.10\pm 0.53$ & 0 & $1.67\pm 0.34$ & $0.04\pm 0.16$ \\
 \enddata
\tablecomments{Molybdenum isotope measurements \citep{dauphas01}. $f_{Re/Mo}$ is the Re/Mo fractionation factor, $\epsilon ^i$ represents the molybdenum isotopic composition after internal normalization, and $\delta$ is the composition after external normalization of raw ${\rm ^{98}Mo/^{96}Mo}$ ratios. All uncertainties are $2\sigma$.}
\end{deluxetable}


\begin{thebibliography}{}

\bibitem[All\`egre et al.(1995)]{allegre95} All\`egre, C.-J., Manh\`es, G., \& G\"opel, C. 1995, Geochim. Cosmochim. Acta, 59, 1445
\bibitem[Anders and Grevesse(1989)]{anders89} Anders, E., and Grevesse, N. 1989, Geochim. Cosmochim. Acta, 53, 197
\bibitem[Anders and Zinner(1993)]{anders93} Anders, E., and Zinner, E. 1993, Meteoritics, 28, 490
\bibitem[Arlandini et al.(1999)]{arlandini99} Arlandini, C., K\"appeler, F., Wisshak, K., Gallino, R., Lugaro, M., Busso, M., and Straniero, O. 1999, Astrophys. J., 525, 886
\bibitem[Becker et al.(1987)]{becker87} Becker, R., Koller, P., Morschl, P., Kiesl, and F. Hermann, F. 1987, in The Allende Meteorite Reference Sample, Smithsonian Contributions to the Earth Sciences 27, ed. E. Jarosewich, R.S. Clarke Jr., and J.N. Barrows (Washington: Smithsonian Institution Press), 16
\bibitem[Burbidge et al.(1957)]{burbidge57} Burbidge, E.M., Burbidge, G.R., Fowler, W.A., and Hoyle, F. 1957, Rev. Mod. Phys., 29, 547
\bibitem[Cameron(1957)]{cameron57} Cameron, A.G.W. 1957, Publ. Astron. Soc. Pacific, 69, 201
\bibitem[Chen and Wasserburg(1996)]{chen96} Chen, J.H., and Wasserburg, G.J. 1996, in Earth Processes: Reading the Isotopic Code, Geophysical Monograph 95, ed. A. Basu, and S.R. Hart (Washington: American Geophysical Union), 1
\bibitem[Clayton(1982)]{clayton82} Clayton, D.D. 1982, Quart. J. Roy. Astron. Soc., 23, 174
\bibitem[Clayton(1993)]{clayton93} Clayton, R.N. 1993, Ann. Rev. Earth Planet. Sci., 21, 115
\bibitem[Dauphas et al.(2001)]{dauphas01} Dauphas, N., Reisberg, L., and Marty, B. 2001, Anal. Chem., 73, 2613
\bibitem[Farquhar et al.(2000)]{farquhar00} Farquhar, J., Savarino, J., Jackson, T.L., and Thiemens, M.H. 2000, Nature, 404, 50
\bibitem[Fegley et al.(1993)]{fegley93} Fegley Jr., B., Lodders, K., and Palme, H. 1993, Meteoritics, 28, 346
\bibitem[Fleet et al.(1999)]{fleet99} Fleet, M.E., Liu, M., and Crocket, J.H. 1999, Geochim. Cosmochim. Acta, 63, 2611
\bibitem[Galer and Goldstein(1996)]{galer96} Galer, S.J.G., and Goldstein, S.L. 1996, in Earth Processes: Reading the Isotopic Code, Geophysical Monograph 95, ed. A. Basu, and S.R. Hart (Washington: American Geophysical Union), 75
\bibitem[Goldstein and Ogilvie(1965)]{goldstein65} Goldstein, J.I., and Ogilvie, R.E. 1965, Geochim. Cosmochim. Acta, 29, 893
\bibitem[Halliday et al.(1996)]{halliday96} Halliday, A., Rehk\"amper, M., Lee, D.-C., and Yi, W. 1996, Earth Planet. Sci. Lett., 142, 75
\bibitem[Harper et al.(1992)]{harper92} Harper Jr., C.L., Weismann, H., and Nyquist, L.E. 1992, Meteoritics, 27, 230
\bibitem[Huang and Masuda(1997)]{huang97}Huang, M., and Masuda, A. 1997, Anal. Chem., 69, 1135
\bibitem[Huss and Lewis(1995)]{huss95}Huss, G.R., and Lewis, R.S. 1995, Geochim. Cosmochim. Acta, 59, 115
\bibitem[Hutcheon et al.(1992)]{hutcheon92} Hutcheon, I.D., Olsen, E., Zipfel, J., and Wasserburg, G.J. 1992, Lunar Planet. Sci, XXIII, 565
\bibitem[Jacobsen and Wasserburg(1984)]{jacobsen84} Jacobsen, S.B., and Wasserburg, G.J. 1984, Earth Planet. Sci. Lett., 67, 137
\bibitem[Lee and Halliday(1995)]{lee95} Lee, D.-C., and Halliday, A.N. 1995, Nature, 378, 771
\bibitem[Lee and Halliday(2000)]{lee00} Lee, D.-C., and Halliday, A.N. 2000, Science, 288, 1629
\bibitem[Liu and Fleet(2001)]{liu01} Liu, M., and Fleet, M.E. 2001, Geochim. Cosmochim. Acta, 65, 671
\bibitem[Lugmair and Shukolyukov(1998)]{lugmair98} Lugmair, G.W., and Shukolyukov, A. 1998, Geochim. Cosmochim. Acta, 62, 2863
\bibitem[Manuel et al.(1972)]{manuel72} Manuel, O.K., Hennecke, E.W., and Sabu, D.D. 1972, Nature, 240, 99
\bibitem[Mar\'echal et al.(1999)]{marechal99} Mar\'echal, C.N., T\'elouk, P., Albar\`ede, F. 1999, Chem. Geol., 156, 251
\bibitem[Mittlefehldt et al.(1998)]{mittlefehldt98} Mittlefehldt, D.W., McCoy, T.J., Goodrich, C.A.,  and Kracher, A. 1998,  in Planetary Materials, Rev. Mineral. 36, ed. J.J. Papike (Washington: Mineralogical Society of America), 4-1
\bibitem[Nicolussi et al.(1998a)]{nicolussi98a} Nicolussi, G.K., Pellin, M.J., Lewis, R.S., Davis, A.M., Amari, S., and Clayton, R.N. 1998a, Geochim. Cosmochim. Acta, 62, 1093 
\bibitem[Nicolussi et al.(1998b)]{nicolussi98b} Nicolussi, G.K., Pellin, M.J., Lewis, R.S., Davis, A.M., Clayton, R.N., and Amari, S. 1998b, ApJ, 504, 492
\bibitem[Niemeyer(1988)]{niemeyer88} Niemeyer, S. 1988, Geochim. Cosmochim. Acta, 52, 2941
\bibitem[Nomura et al.(1983)]{nomura83}Nomura, M., Kogure, K., and Okamoto, M. 1983, Int. J. Mass Spectrom. Ion Phys., 50, 219
\bibitem[Pellin et al.(1999)]{pellin99} Pellin, M.J., Davis, A.M., Lewis, R.S., Amari, S., and Clayton, R.N. 1999, Lunar Planet. Sci., XXX, \# 1969 
\bibitem[Pellin et al.(2000)]{pellin00} Pellin, M.J., Davis, A.M., Calaway, W.F., Lewis, R.S., Clayton, R.N., and Amari, S. 2000, Lunar Planet. Sci., XXXI, \# 1934
\bibitem[Podosek and Nichols(1996)]{podosek96} Podosek, F.A., and Nichols Jr., R.H. 1996, in Astrophysical Implications of the Laboratory Study of Presolar Materials, ed. T.J. Bernatowicz and E.K. Zinner (Woodbury: American Institute of Physics), 617
\bibitem[Podosek et al.(1999)]{podosek99} Podosek, F.A., Nichols Jr., R.H., Brannon, J.C., and Dougherty, J.R. 1999, Lunar Planet. Sci.,  XXX, \# 1307
\bibitem[Qi-Lu and Masuda(1992a)]{qilu92a}Qi-Lu, and Masuda, A. 1992a, J. Am. Soc. Mass Spectrom.,3, 10
\bibitem[Qi-Lu and Masuda(1992b)]{qilu92b}Qi-Lu, and Masuda, A. 1992b, Analyst, 117, 869
\bibitem[Qi-Lu and Masuda(1994)]{qilu94}Qi-Lu, and Masuda, A. 1994, Int. J. Mass Spectrom. Ion Proc., 130, 65
\bibitem[Reynolds and Turner(1964)]{reynolds64} Reynolds, J.H., and Turner, G. 1964, J. Geophys. Res., 69, 3263 
\bibitem[Righter and Drake(1997)]{righter97} Righter, K., and Drake, M.J. 1997, Earth Planet. Sci. Lett., 146, 541
\bibitem[Schmitt et al.(1989)]{schmitt89}  Schmitt, W., Palme, H., and W\"anke, H. 1989, Geochim. Cosmochim. Acta, 53, 173
\bibitem[Scott and Wasson(1975)]{scott75} Scott, E.R.D., and Wasson, J.T. 1975, Rev. Geophys. Space Phys., 13, 527
\bibitem[Sears and Dodd(1988)]{sears88} Sears, D.W.G., and Dodd, R.T. 1988, in Meteorites and the Early Solar System, ed. J.F. Kerridge, and  M.S. Matthews (Tucson: University of Arizona Press), 3
\bibitem[Shen et al.(1996)]{shen96} Shen, J.J., Papanastassiou, D.A., and Wasserburg, G.J. 1996, Geochim. Cosmochim. Acta, 60, 2887
\bibitem[Shukolyukov and Lugmair(1998)]{shukolyukov98} Shukolyukov, A., and Lugmair, G.W. 1998, Science, 282, 927 
\bibitem[Shukolyukov and Lugmair(2001)]{shukolyukov01} Shukolyukov, A., and Lugmair, G.W. 2001, Lunar Planet. Sci., XXXII, 1365
\bibitem[Smoliar et al.(1996)]{smoliar96} Smoliar, M.I., Walker, R.J., and Morgan, J.W. 1996, Science, 271, 1099
\bibitem[Thiemens(1999)]{thiemens99} Thiemens, M.H. 1999, Science, 283, 341
\bibitem[Thiemens and Jackson(1995)]{thiemens95} Thiemens, M.H., and Jackson, T. 1995, Lunar Planet. Sci., XXVI, 1405
\bibitem[Wood(1964)]{wood64} Wood, J.A. 1964, Icarus, 3, 429 
\bibitem[Yin and Jacobsen(1998)]{yin98} Yin, Q.Z., and Jacobsen, S.B. 1998, Lunar Planet. Sci., XXIX, \# 1802
\bibitem[Yin et al.(2000)]{yin00} Yin, Q.Z., Yamashita, K., and Jacobsen, S.B. 2000, Lunar Planet. Sci., XXXI, \# 1920
\bibitem[Yin and Jacobsen(2000)]{yin00b} Yin, Q.Z., and Jacobsen, S.B. 2000, Meteoritics Planet. Sci., 35, A175
\bibitem[Yin et al.(2001)]{yin01} Yin, Q., Jacobsen, S.B., Blichert-Toft, J., T\'elouk, P., and Albar\`ede, F. 2001, Lunar Planet. Sci., XXXII, \# 2128
\end{thebibliography}
\end{document}